\documentclass[aps,prl,a4paper,twocolumn,showpacs,showkeys,preprintnumbers,amsmath,amssymb]{revtex4}
\usepackage{graphicx}
\usepackage{bm}
\usepackage{dcolumn}
\usepackage{color}
\usepackage{pst-plot}
\renewcommand{\texttt}{{}}
\newcommand{\be}{\begin{eqnarray}}
\newcommand{\ee}{\end{eqnarray}}
\usepackage{graphicx}
\begin{document}
%\title{The microstructure of a quantum spacetime}
\title{Fuzziness at the horizon} % of Loop Quantum Gravity}
\author{Davide Batic}
\thanks{Electronic address: dbatic@uniandes.edu.co}
\affiliation{Departamento de Matematica, Universidad de los Andes, %% PRD VERSION
Cra 1E, No. 18A-10,                                                %% PRD VERSION
Bogot\'a,                                                          %% PRD VERSION
Colombia \\Department of Mathematics, University of West Indies,
Kingston, Jamaica}

\author{Piero Nicolini}
\thanks{Electronic address: nicolini@th.physik.uni-frankfurt.de}
\affiliation{Frankfurt Institute for Advanced Studies (FIAS),
%Johann Wolfgang Goethe University
Institut f\"ur Theoretische Physik, Johann Wolfgang Goethe-Universit\"at, Ruth-Moufang-Strasse 1,
60438 Frankfurt am Main, Germany}

\date{\small\today}
%\title                                                                %% ARXIV VERSION
%{{\bf On the Cauchy horizon of the noncommutative geometry            %% ARXIV VERSION
%inspired black hole}}                                                   %% ARXIV VERSION

%% \author                                                            %% PRD VERSION
%% {Davide Batic$^1$}                                                 %% PRD VERSION
%% \email{dbatic@uniandes.edu.co}                                     %% PRD VERSION

%% \author{Piero Nicolini$^2$}                                        %% PRD VERSION
%% \email{nicolini@th.physik.uni-frankfurt.de}                        %% PRD VERSION
%% \affiliation{$^1$Departamento de Matematica, Universidad de los Andes, %% PRD VERSION
%% Cra 1E, No. 18A-10,                                                %% PRD VERSION
%% Bogot\'a,                                                          %% PRD VERSION
%% Colombia\\
%% $^2$Institut f\"{u}r Theoretische Physik, Johann Wolfgang Goethe Universit\"{a}t,
 %% PRD VERSION
%M.von Laue Str.1,                                                %% PRD VERSION
%% 60438 Frankfurt am Main, Germany}                                                          %% PRD VERSION
                                                      %% PRD VERSION
%$^2$Inst. f\"{u}r Theoretische Physik, J.W. Goethe-Universit\"{a}t, Max-von-Laue-Str.1, 60438 Frankfurt am Main, Germany  }                                         %% PRD VERSION
                                                                      %% PRD VERSION
                                                                     %% PRD VERSION

%J.W.
%}                                                                    %% PRD VERSION

\date{\today}% It is always \today, today,
%\documentclass[published]{JHEP3} % 10pt is ignored!

%\JHEP{00(2009)000}

%\JHEPspecialurl{http://jhep.sissa.it/JOURNAL/JHEP3.tar.gz}

%\usepackage{epsfig,multicol,bbm}

%Begin special definitions for Instructions file
%\newcommand{\ttbs}{\char'134}%\backslash for \tt
%\newcommand\fverb{\setbox\fverbbox=\hbox\bgroup\verb}
%\newcommand\fverbdo{\egroup\medskip\noindent%
%           \fbox{\unhbox\fverbbox}\ }
%\newcommand\fverbit{\egroup\item[\fbox{\unhbox\fverbbox}]}
%\newbox\fverbbox
%\newcommand{\jhepname}{JHEP}
%end

%\preprint{\hepth{\cdots}}  % OR: \preprint{Aaaa/Mm/Yy\\Aaa-aa/Nnnnnn}
                % Use \hepth etc. also in bibliography.
\begin{abstract}
We study the stability of the noncommutative Schwarzschild black hole interior
 by analysing the propagation of a massless scalar field between the two horizons.
We show that the spacetime fuzziness triggered by the field higher momenta
can cure the classical exponential blue-shift      %% PRD VERSION
divergence, suppressing the emergence of infinite energy                                %% PRD VERSION
density in a region nearby the Cauchy horizon.                                        %% PRD VERSION
\end{abstract}

%\dedicated{Dedicated to\ldots\\if you want.}

\pacs{Valid PACS appear here}% PACS, the Physics and Astronomy         %% PRD VERSION
%Use showkeys class option if keyword
\maketitle

Black holes can exhibit not only an event horizon, namely the outermost surface that physically separates two noncommunicating regions of spacetime, but also inner Cauchy horizons. These internal horizons, null surfaces beyond which spacetime predictability breaks down,  have the intriguing properties of showing up a ``dual effect'' of the conventional red shift, i.e. the blue shift. 
To understand the physics of this blue shift, it is common procedure to study a radiation represented by a scalar field, propagating in the region between the two horizons. For the sake of clarity we consider a spherically symmetric spacetime region, whose metric can be cast in the form
\begin{equation}
ds^2=\frac{r^2dr^2}{(r_+-r)(r-r_-)}-\frac{(r_+-r)(r-r_-)}{r^2}dt^2-r^2d\Omega^2
\end{equation} 
where $r_-$ is the Cauchy horizon, $r_+$ is the event horizon, i.e. $r_-<r<r_+$, $r$ plays the role of a temporal coordinate and $t$  is a spatial one. Introducing tortoise coordinates 
\begin{equation}
r^\star=-r-\frac{1}{\kappa_+}\ln(r_+-r)+\frac{1}{\kappa_-}\ln(r-r_-)
\end{equation}
where $\kappa_\pm\equiv (r_+-r_-)/r_\pm^2$, we can define null coordinates $x_-=-r^\ast-t$ and $x_+=-r^\ast+t$ to study the propagation of a scalar field on this background in terms of the scalar wave equation
\begin{equation}\label{KG}
g^{\mu\nu}\nabla_\mu\nabla_\nu\phi=0.
\end{equation}
The solution of the above equation, let us conclude that the field,  in the vicinity of the Cauchy horizon where $x_+\to \infty$, decays as $\phi\sim x_+^{-2\ell-2}$, with $\ell$ the multipole order of the field. However this is no longer true for the energy of the field. Indeed, if we consider, the field's rate variation as measured by a free falling observer (FFO) crossing the Cauchy horizon we obtain the infinite result
\begin{equation}
\phi_{,\mu}\ U^\mu\simeq\phi_{,x_+}\ \dot{x}_+\ \sim \  x_+^{-2\ell-3}\ e^{\kappa_-x_+}
\label{FFOdiv}
\end{equation}
where $U^\mu$ is the 4-velocity of the observer (the dot denotes differentiation with respect
to proper time). Actually the FFO measures a flux of energy given by the square of the above quantity, that is even more divergent. This mechanism of instability due to the infinite blue shift at the Cauchy horizon can be explained in these terms. An external observer would require an infinite time to reach the future null infinity ($x_+=\infty$) since at the best its velocity is $\dot{x}_+\simeq 1$. On the other hand, a FFO can reach the Cauchy horizon in a finite proper time, which implies that $\dot{x}_+$ will diverge as $x\to\infty$. From Eq. (\ref{FFOdiv}) we see that this divergence overcomes the field decay. As a result the Cauchy horizon is unstable.
%In other words a free falling observer (FFO) would experience that any incoming radiation piles up an infinite amount of energy density, disrupting the spacetime geometry and developing unbounded curvatures.
%There is a lot of cases in which Cauchy horizons might destabilize the spacetime geometry, for instance the Reisser-Nordstr\"{o}m, Kerr and Kerr-Newman inner horizons or the white hole anti horizon. 
An extensive study on this subject has basically led to the general conclusion that this infinite amount of energy density at the Cauchy horizon can develop unbounded curvature, disrupting the spacetime geometry
\cite{Simpson:1973ua,Eardley:1974zz,McNamara:1979a,McNamara:1979b,Gursel:1979zza,Gursel:1979zz,Matzner:1979zz,Poisson:1989zz,Poisson:1990eh,Brady:1992,Burko:1995uq,Markovic:1994gy,Balbinot:1989,Balbinot:1991,Balbinot:1993rf,Konoplya:2008au,Cardoso:2010rz,Dotti:2010uc}.

Up to now we have mentioned purely classical solutions, since the above analyses have concerned quantum effect at the most in the matter fields propagating on the spacetime manifold. On the other hand, there is a recent class of black hole solutions (QGBHs) obtained by means of quantum gravity arguments, like loop quantum black holes \cite{Modesto:2006mx,Modesto:2008im,Modesto:2009ve}, asymptotically safe gravity black holes \cite{Bonanno:2000ep,Bonanno:2006eu}, generalized uncertainty principle \cite{AmelinoCamelia:2005ik,Husain:2008qx} and noncommutative geometry inspired black holes \cite{Nicolini:2005zi,Nicolini:2005vd,Rizzo:2006zb,Ansoldi:2006vg,Spallucci:2008ez,Arraut:2009an,Nicolini:2009gw}
 (for a review see \cite{Nicolini:2008aj} and
the references therein). Independently on their starting point the above solutions, converge on a unique qualitative behavior, namely the absence of any curvature singularity
% a positive heat capacity thermodynamically stable final stage of the Hawking evaporation
and the presence of more than a horizon. In other words, as far as some sort of smearing effect is concerned due to the fuzziness of spacetime in its quantum gravity regime, the physics of QGBHs has a universal character. This fact has its equivalent on the thermodynamics side: the Hawking temperature admits a maximum, followed by a ``SCRAM phase'', a thermodynamic stable  shut down, characterized by a positive black hole heat capacity.  As a consequence, also for the neutral solution, in place of the runaway behavior of the temperature, one finds that the evaporation ends up with a zero temperature
extremal black hole, a final configuration entirely governed by a quantum gravity induced minimal length. This new scenario of the evaporation implies a further virtue of QGBHs: a finite temperature prevents any relevant back reaction, namely a self interaction of the radiated energy with its source. Thus we can
conclude that these solutions are stable versus back reaction and can describe the entire black hole life until the
final configuration.  However, having an inner Cauchy horizon is again a source of concern, since we might have the suspect that
the interior of these black holes is unstable. As a result the solutions could be no longer singularity free, since the
singularity might occur on the inner horizon rather than at the origin, frustrating the efforts that vivified the formulations at the basis of their derivation.  In some sense, the instability of QGBHs is even worse with respect to the conventional classical analogs, since it affects the neutral, static case too. 
%Furthermore, the concrete possibility of producing mini black holes in particle detectors and observing the Hawking radiation according to the scenario provided for QGBHs makes the situation even more puzzling and therefore a deeper understanding of the destiny of their interiors is more urgent than ever.
To address this problem we need to change our perspective. If we do believe in the tenets of quantum gravity, we
have to accept the possibility for a quantum manifold to provide a natural ultraviolet cut-off for any field propagating over it in order to prevent any growth of energy beyond Planckian magnitude.  
%As a starting point, we will consider in this paper a classical perturbation of the manifold represented by a scalar field crossing the event horizon and approaching the inner Cauchy horizon. 
Without loss of generality we will consider the neutral noncommutative spacetime only, even if our analysis holds for the charged solution too and other QGBHs. Indeed noncommutative geometry is just one of the possible effective ways to implement a natural cut-off.

We shall start recalling some properties of the noncommutative geometry inspired black hole, whose line element is given in
\cite{Nicolini:2005vd}
\begin{equation}\label{NCSS}
ds^2=\left(1-2m(r)/r\right)dt^2
-\frac{dr^2}{\left(1-2m(r)/r\right)}-r^2d\Omega^2
\end{equation}
with the mass function $m(r)=4M\gamma\left(3/2, r^2/4 \theta \right)/ \sqrt{\pi}$, where $M$ is the total mass in
the spacetime manifold, $\theta$ is a parameter encoding noncommutativity and having the dimension of a length
squared, while 
\begin{equation}
\gamma\left(3/2, r^2/4 \theta \right)=\int_0^{r^2/4\theta}dt\ t^{1/2}\ e^{-t}
\end{equation} 
is the incomplete lower gamma function. The above line element is clearly regular at the
origin, where a de Sitter core accounts for the mean value of the quantum fluctuations of the manifold.  The metric
admits one, two or no horizon depending on the total mass $M$, respectively equal, larger or smaller than extremal
black hole total mass $M_0\approx 1.9\sqrt{\theta}$. In the following we shall restrict our attention to the case
of two horizons%, an (outer) event horizon  $r_+$ and a (inner) Cauchy horizon $r_-$
. In this scenario the line
element (\ref{NCSS}) in the interior region $r_{-}<r<r_{+}$ can be cast in the form
\begin{equation}\label{metric2}
ds^2=\frac{d\tau^2}{\left(\frac{2m(\tau)}{\tau}-1\right)}
-\left(\frac{2m(\tau)}{\tau}-1\right)d\rho^2-\tau^2d\Omega^2
\end{equation}
where we introduced the new variables $\tau$ and $\rho$ in place
of $r$ and $t$, since in this region they become a temporal and a
spatial coordinate, respectively. It is convenient to introduce a
temporal tortoise coordinate $\tau^{*}$ defined as
$d\tau^{*}=d\tau/\left(2 m(\tau)/\tau-1\right)$ such that
$\tau^{*}\to\pm\infty$ as $\tau\to r_{\pm}$. In the sequel we
shall introduce null coordinates $x_{-}\equiv-\tau^{*}-\rho$ and
$x_{+}\equiv-\tau^{*}+\rho$ as in
\cite{Gursel:1979zza,Gursel:1979zz}. Then the metric reads
\begin{equation}\label{metric3}
ds^2={\left(\frac{2m(\tau)}{\tau}-1\right)}dx_{-}dx_{+}-\tau^2d\Omega^2.
\end{equation}
The event horizon becomes the null hypersurface $x_{-}=-\infty$
and the left and right branches of the Cauchy horizon $r_{-}$ are
null hypersurfaces $x_{-}=\infty$ and $x_{+}=\infty$,
respectively. In the region between the two horizons (see Fig. 1) we consider
the propagation of a massless scalar test field
$\phi=\phi(\tau,\rho,\vartheta,\varphi)$ governed by the equation (\ref{KG})
where the metric is associated to the background geometry (\ref{metric2}).
Now it is the time to invoke the noncommutative nature of the field. Indeed, up to now, noncommutative
effects have been considered to smear the matter generating the background geometry only. We need to
extend this procedure to matter propagating over the manifold too. To this purpose, we follow the formulation
proposed in
\cite{Cho:1999sg,Smailagic:2003yb,Smailagic:2003rp,Smailagic:2004yy,Spallucci:2006zj,Banerjee:2009gr,Banerjee:2009xx}
%Indeed the manifold fluctuates and the very concept of point is little meaningful.
to get a modified integral measure in the momentum representation
of the field, which between the two horizons can be written as
\begin{eqnarray}
&&\phi (\tau^{*},\rho,\vartheta,\varphi)=\\
&&=\sum_{\ell=0}^{\infty}\sum_{m=-\ell}^{\ell}\int_{-\infty}^{\infty}dk e^{-k^2\theta/4}e^{-ik\rho}\ \displaystyle{
\frac{1}{\tau}}\ \psi_{\ell mk}(\tau^{*}) Y_{\ell m}.\nonumber
\end{eqnarray}
The presence of an exponential damping factor encodes the effect of the noncommutative UV regularization.
Here $\tau$ has to be thought as an implicit function of the variable $\tau^{*}$, while
$Y_{\ell m}=Y_{\ell m}(\vartheta,\varphi)$
denotes the spherical harmonics.
Analogous modifications have been already efficiently employed in a variety of contexts,
namely to describe a traversable wormhole sustained by quantum geometry fluctuations \cite{Garattini:2008xz},
to remove the initial cosmological singularity and drive the inflation without an inflaton field \cite{Rinaldi:2009ba}
and to get corrections to the Unruh thermal bath by means of a nonlocal deformation of conventional
field theories \cite{Casadio:2005vg,Casadio:2007ec,Nicolaevici:2008zz,Nicolini:2009dr}.
Further contributions concern the modification of the Newton potential in the presence of
noncommutative spacetime coordinates \cite{Gruppuso:2005yw}, the evaporation of the noncommutative
black holes in terms of gravitational amplitudes for boson and fermion fields
\cite{DiGrezia:2006rw,DiGrezia:2007gr,Di Grezia:2007uy} and for up to ten spatial
dimensions in particle detectors at the LHC \cite{Casadio:2008qy} and the
calculation of the spectral dimension of a quantum spacetime \cite{Modesto:2009qc}.
 As a result we obtain the ``radial'' function $\psi_{\ell mk}$ obeys the equation
\begin{equation}\label{radial}
\frac{d^2\psi_{\ell mk}}{d{\tau^{*}}^2}+\left[k^2-V(\tau^{*})\right]\psi=0
\end{equation}
where the potential is given by
\begin{equation}
V(\tau^{*})=\frac{g_{\rho\rho}}{\tau}\left[\frac{\ell(\ell+1)}{\tau}+2\frac{g_{\rho\rho}}{\tau}+\partial_\tau g_{\rho\rho}\right].
\end{equation}
For our purposes we are interested in the asymptotic solutions of (\ref{radial}) as
$\tau^{*}\to-\infty$. The potential $V(\tau^{*})$ decays exponentially in time as
\begin{equation}
V(\tau^{*})\approx e^{\mp\alpha_{\pm}\tau^{*}},\qquad \tau^{*}\to\pm\infty.
\end{equation}
where $2\alpha_{\pm}\equiv\pm (dg_{00}/dr)_{r=r_{\pm}}$. Finally,
near the Cauchy horizon the asymptotic solutions of (\ref{radial})
are
\begin{equation}
e^{-ik\rho}\psi_{\ell mk}(\tau^{*})\approx\ e^{\pm ikx_{\pm}}\ \left[1+O(e^{\alpha_{-}\tau^{*}})\right].
\end{equation}
Up to exponentially vanishing corrections, the solution are plane waves approaching the left and right branch of
the horizon $r_{-}$, respectively.  As in \cite{Matzner:1979zz} the energy density in the scalar field $\phi$ as
measured by a freely falling observer near a horizon with four velocity $U^{\mu}$ will be proportional to
\begin{equation}
{\cal E}=(\phi_{,~\alpha}U^\alpha)(\phi_{,~\beta}U^\beta)+\frac{1}{2}\phi_{,~\alpha}{\phi^{*}}^{,~\alpha}.
\end{equation}
Since $-\tau^{*}\pm \rho=const$ are null surfaces and taking into account for the form of $\phi$ nearby the horizon,
 we have that the energy density is dominated by the term $|\phi_{,~\alpha}U^\alpha|^2$. Therefore we can
 restrict our analysis to the term $\phi_{,~\alpha}U^\alpha$ only. To this purpose, we need the form of the
 velocity vector field associated to the FFO which can be written as
\begin{equation}
U=U^{x_{-}}\frac{\partial}{\partial
x_{-}}+U^{x_{+}}\frac{\partial}{\partial x_{+}}
\end{equation}
%\begin{equation}
%U=(U^\rho-U^{\tau^{*}})\frac{\partial}{\partial v}+(-U^\rho-U^{\tau^{*}})\frac{\partial}{\partial u}
%\end{equation}
where
%\footnote{The sign of the component $U^{\tau^{*}}$ changes at the turning points $r_{+}$ and $r_{-}$.
%Since $r$ is a timelike coordinate in the inner region ($III$) between the two horizons we shall take the
%negative sign in the expression for $U^{\tau^{*}}$. }
%\be
%U^\rho&=&-\frac{h\tau}{2m(\tau)-\tau} \\ U^{\tau^{*}}&=&\ \mp \ U^\rho\sqrt{1+\frac{2m(\tau)-\tau}{h^2\tau}}.
%\ee
%with $h$ a dimensionless parameter.
%Thus, we conclude that
\begin{equation}
U^{x_{\pm}}=-\frac{\tau}{2m(\tau)-\tau}\left(h\mp\sqrt{h^2+\frac{2m(\tau)-\tau}{\tau}}\right),
\end{equation}
with $h$ a dimensionless parameter. If $h>0$ the FFO worldline
enters region $III$ from region $I$  and exits region $III$
through the left-hand branch ($x_{-}=\infty$) of the inner
horizon. If $h<0$ the worldline enters region $III$ from region
$II$ and exits through the right-hand ($x_{+}=\infty$) branch of
$r_{-}$. If $h=0$ the worldline will move through the region $III$
passing through the bifurcation points of the horizon $r_{-}$. In
the vicinity of the Cauchy horizon, we have for $h>0$
\begin{equation}
U^{x_{+}}\approx \frac{1}{2},\quad U^{x_{-}}\approx
e^{-\alpha_{-}\tau^{*}}=e^{\alpha_{-}(x_{-}+x_{+})/2}
\end{equation}
whereas for $h<0$
\begin{equation}
U^{x_{+}}\approx e^{\alpha_{-}(x_{-}+x_{+})/2},\quad
U^{x_{-}}\approx\frac{1}{2}.
\end{equation}
Hence, for $h>0$ and asymptotically for $x_{-}\to\infty$ we have
\begin{equation}\label{A}
\phi_{,~\alpha}U^\alpha\approx
e^{\alpha_{-}(x_{-}+x_{+})/2}\frac{\partial\phi}{\partial
x_{-}}+\frac{1}{2}\frac{\partial\phi}{\partial x_{+}}
\end{equation}
whereas for $h<0$ and $x_{+}\to\infty$
\begin{equation}\label{B}
\phi_{,~\alpha}U^\alpha\approx
\frac{1}{2}\frac{\partial\phi}{\partial x_{-}}+
e^{\alpha_{-}(x_{-}+x_{+})/2}\frac{\partial\phi}{\partial x_{+}}.
\end{equation}
In order that the FFO can measure a nondivergent amount of field
energy density near the $r_{-}$ horizon, the appropriate
derivative of the field times the exponential blue-shift factor
must be finite. In the classical picture
\cite{Gursel:1979zza,Matzner:1979zz} from the last two relations
above one concludes that the $e^{-ikx_{+}}$ waves are singular
along the left branch of $r^{-}$ and the $e^{ikx_{-}}$ waves
become singular along the right branch of $r_{-}$. We shall see
that due to the noncommutativity of the field, (\ref{A}) and
(\ref{B}) stay bounded at the Cauchy horizon. Indeed, we find that
for the left-going component
\begin{equation}\label{C}
\phi_{,~\alpha}U^\alpha\sim \frac{x_{-}}{\theta^{3/2}}\
e^{\alpha_{-}(x_{-}+x_{+})/2}e^{-x_{-}^2/\theta}
\end{equation}
which vanishes as $x_{-}\to\infty$, keeping $x_{+}$ constant.
Analogously, we find for the right-going component
\begin{equation}
\phi_{,~\alpha}U^\alpha\sim \frac{x_{+}}{\theta^{3/2}}\
e^{\alpha_{-}(x_{-}+x_{+})/2}\ e^{-x_{+}^2/\theta}
\end{equation}
which vanishes as $x_{+}\to\infty$, for $x_{-}$ fixed. The above
result confirms that, probing higher momenta the field basically
triggers the noncommutative nature of the manifold, which shows
graininess and prevents any spacetime resolution beyond the value
$\sqrt\theta$. This let us also conclude that in this framework no
mass inflation can occur. To this purpose, we define \be
T_{ab}={\cal E}_{in}l_al_b+{\cal E}_{out}n_an_b \ee as the two-dimensional section of the stress tensor, which describes the
cross flowing stream of infalling and outgoing of light like
particles following null geodesics. Here $l_a$ is the radial null
vector pointing inwards, $n_a$ is the radial null vector pointing
outwards, while ${\cal E}_{in}$ and ${\cal E}_{out}$ represent the
energy density of the fluxes. The mass inflation is a huge boom of
the black hole internal mass parameter, which becomes classically
unbounded at the Cauchy horizon. Contrary to the intuition, the
inflation is due to both the outflux and the blueshifted influx of
a collapsing star as shown in \cite{Poisson:1990eh}. On the other
hand, in the present framework, energy densities cannot diverge
even at the Cauchy horizon. Therefore, the mass inflation which is
in general proportional to the product $T^{ab}T_{ab}$ will not
take place. The above analysis concerns the leading contribution
to the energy density of the field in the vicinity of the Cauchy
horizon along the lines of \cite{Gursel:1979zza,Gursel:1979zz}.
Since higher order terms fall off faster, we argue that the
stability of QGBH interiors can be shown in general. However,
according to the theory of the stability in
\cite{McNamara:1979a,McNamara:1979b}, an exponential decay is a
mere necessary condition only: in other words even if each term of
the expansion is vanishing, their global contribution could yet
destabilize the solution. Furthermore, we have addressed here the
most simple case of classical perturbation of the manifold, while
in general the field could be quantized. In such a case, according
to previous contributions \cite{Balbinot:1993rf}, the stress
tensor $\left<T_{ab}\right>$ is expected to have an even worse UV
behavior. For the above reasons, we think that, after the present
analysis, further investigations will be necessary, also to
include the other spacetimes within the class of QGBHs.

\begin{figure}
\begin{center}
\includegraphics[height=7.5cm]{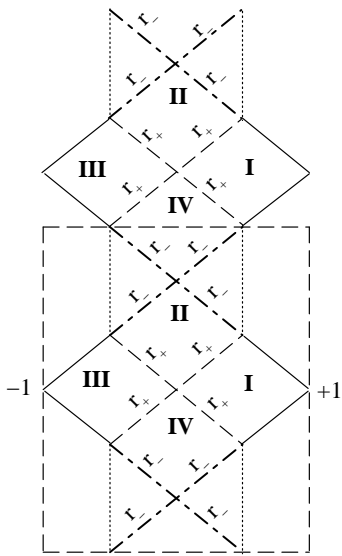}
\caption{\label{foot}The Carter-Penrose diagram of the manifold. The conformal diagram of the maximally extended noncommutative inspired
Schwarzschild spacetime. The radii $r_{\pm}$ represent the outer and inner horizons,
respectively. The central singularity appearing in the Reissner-Nordstr\"{o}m metric
is now replaced by a regular de Sitter core, dotted line. The upper and bottom part
of the box indicated by the dashed line can be identified to make the manifold
cyclic in the time coordinate.
}
\end{center}
\end{figure}

%% \begin{figure}
%% \begin{center}
%% \includegraphics[height=7.5cm]{PenroseNicolini3.eps}
%% \caption{\label{foot}The Carter-Penrsore diagram of the manifold. The conformal diagram of the maximally extended noncommutative inspired
%% Schwarzschild spacetime. $r_{\pm}$ represent the outer and inner horizons,
%% respectively. The central singularity appearing in the Reissner-Nordstr\"{o}m metric
%% is now replaced by a regular deSitter core, dotted line. The upper and bottom part
%% of the box indicated by the dashed line can be identified to make the manifold
%% cyclic in the time coordinate.
%% }
%% \end{center}
%% \end{figure}

\subsection*{Acknowledgments}
P.N. would like to thank the Universidad de los Andes, Bogot\'a, Colombia for the kind hospitality during the period of work on this project.
P.N. is supported by the Helmholtz International Center for FAIR within the
framework of the LOEWE program (Landesoffensive zur EntwicklungWissenschaftlich-\"{O}konomischer
Exzellenz) launched by the State of Hesse.
%and Roberto Balbinot and Euro Spallucci for valuable hints and comments.

\end{document}